\documentclass[preprint,prd]{revtex4}%
\usepackage{amsmath}
\usepackage{graphicx}
\usepackage{amsfonts}
\usepackage{amssymb}%
\begin{document}
\title{Low-Energy-Theorem Approach to one-particle singularity in QED$_{2+1}$}
\author{Yuichi Hoshino}
\affiliation{Kushiro National College of Technology,Otanoshike nishi 2-32-1,Kushiro,
Hokkaido 084,Japan}

\begin{abstract}
We evaluate the propagator of scalar and spinor in three dimensional quantum
electrodynamics with the use of Ward-Identity for soft-photon emission
vertex.We work well in position space to treat infrared divergences in our
model. Exponentiation of one-photon matrix element yields a full propagator in
position space.It has a simple form as free propagator multiplied by quantum
correction.And it shows a new type of mass singularity.But this is not an
integrable function so that analysis in momentum space is not easy.Term by
term integral converges and they have a logarithmic singularity associated
with renormalized mass in perturbation theory$.$Renormalization constant
vanishes for weak coupling,which suggests confinement of charged
particle.There exsists a critical coupling constant above which the vacuum
expectation value of pair condensation is finite.

\end{abstract}
\maketitle
\tableofcontents

\section{ Introduction}

In three dimensional gauge theory infrared divergences is severer than that in
four dimension.In 1981,Jackiw first demonstrated it in the massless fermionic
$QED_{2+1}$ and after that other authors introduced the infrared counter terms
to renormalize the infrared divergences[1,2].There had been an attempts to
solve Dyson-Schwinger equations to examine the dynamical mass generation or
chiral symmetry breakings in this model to improve low-energy
behaviour[3,4].Another important feature of this model is that it allows
parity violating Chern-Simons term in the Lagrangean without violating gauge
invariance[5].But it has not yet been clear the quantum effects of
Chern-Simons term. During the same period infrared behaviour of the propagator
in the presence of dynamical mass has been analysed to search the physical cut
associated with massless photon[6,7].To maintain gauge covariance Delbourgo
and Jackiw addded the vertex correction with Ward-Identities for soft
particles in the determination of the infrared behaviour of the propagator
with bare mass[9,10].In their works relation between bare and renormalized
masses are given by the spectral function.Their works are fundamental and
important to understand the effects of soft-photons in comparison with
perturbation theory.Of course their works includes non-perurbative effects.In
1992 Delbourgo applied his method(gauge technique) to $QED_{2+1}$ and shown
that massless loop correction to photon soften infrared divergences as
T.Appelquist et.al[3,6].In reference[9] infrared behaviour of the scalar
propagator in the presence of massless fields (photon, graviton) in four
dimension was determined by solving the spectral function.In this work we
apply the same method in three dimension to determine the infrared behaviour
of the propagator.In section I we analyse scalar QED$_{3}$ and evaluate the
scalar propagator in position space based on spectral representation.In this
case an approximaiton is made to choose the one-photon,one-meson intermediate
state to derive the matrix element $\left\langle \Omega|\phi(x)|n\right\rangle
$.Most singular infrared part is assumed to be the contribution of soft photon
emitted from external lines.Therefore we introduce photon mass to avoid
infrared divergences.Lowest order spectral function contains mass and wave
function renormalization up to logarithm in position space. Exponentiation of
the lowest-order spectral function yields the full propagator in position
space.It shows us a new type of mass singurality. Especially mass
renormalization make a drastic change of the propagator.Its has a simple form
as free propagator (with renormalized mass) multiplied by quantum
correction.However this function is not integrable and it is difficult to make
a fourier transform of it.In the perturbative analysises the propagator have
linear and logarithmic infrared divergences near on-shell.We mention the gauge
transformation property and see our solution satisfies Landau-Kharatonikov
transformation.As far as the renormalization constant is concerned we evaluate
it by position space propagator.The result is $Z=0$ for weak coupling,which is
a signpost of confining phase in our approximation.In section II we study
QED$_{3}$ with spinor and see the spectral function for fermion is the same in
section I.In this case there is a interesting possibility of pair
condensation.Evaluating the vacuum expectation value $\left\langle
\overline{\psi}\psi\right\rangle $ and we find that there exists a critical
coupling constant ($e^{2}/(8\pi m)=1)$ above which $\left\langle
\overline{\psi}\psi\right\rangle $ remains finite.

\section{ Scalar QED}

First we assume the Kallen-Lehmann spectral representation of the propagator
for massive boson[1]:%

\begin{align}
\Delta(p^{2})  &  =\int d^{3}x\exp(ip\cdot x)\left\langle \Omega|T\phi
(x)\phi^{+}(0)|\Omega\right\rangle ,\\
\Delta(p^{2})  &  =\int\frac{\sigma(s)ds}{p^{2}-s+i\epsilon},\nonumber\\
\operatorname{Im}\mathit{\Delta(p)}  &  =\sigma(p^{2})=(2\pi)^{2}\sum
_{N}\delta(p-p_{N})\left\langle \Omega|\phi|N\right\rangle \left\langle
N|\phi^{+}|\Omega\right\rangle ,\\
|N  &  >=|r;k_{1},.....k_{n}>,r^{2}=m^{2}.
\end{align}
The spectral function is formally written as a sum of multi-photon
intermediate states:%
\begin{align}
\sigma(p^{2})  &  =\int\frac{d^{2}r}{2r_{0}}\sum_{n=0}^{\infty}\frac{1}%
{n!}(\int\frac{d^{3}k}{(2\pi)^{2}}\theta(k_{0})\delta^{(3)}(k^{2}%
)\sum_{\epsilon})_{n}\delta(p-r-\sum_{i=1}^{n}k_{i})\nonumber\\
&  \times\left\langle \Omega|\phi|r_{i};k_{1},...k_{n}\right\rangle
\left\langle r;k_{1},..k_{n}|\phi^{+}|\Omega\right\rangle .
\end{align}
Here we use the {\normalsize notation}%
\begin{align}
(f(k))_{0}  &  =1,\nonumber\\
(f(k))_{n}  &  =\prod_{i=1}^{n}f(k_{i}).
\end{align}
To evaluate the contribution of the soft-photons,we consider from the
beginnings when only the $n$th photon is soft.Define the following matrix
element%
\begin{equation}
T_{n}(r;k_{1},..,k_{n})=\left\langle \Omega|\phi|r;k_{1},..,k_{n}\right\rangle
.
\end{equation}
We consider $T_{n}$ for $k_{n}^{2}\neq0,$we continue off the photon mass-shell
by Lehmann-Symanzik-Zimmermann (LSZ) formula:%

\begin{align}
T_{n}  &  =\epsilon_{n}^{\mu}T_{n\mu},\\
T_{n}^{\mu}  &  =\int d^{3}x\exp(ik_{n}\cdot x)\square_{x}\left\langle
\Omega|T\phi A_{\mu}(x)|r;k_{1},..,k_{n}\right\rangle \nonumber\\
&  =\int d^{3}x\exp(ik_{n}\cdot x)\left\langle \Omega|T\phi j^{\mu}%
(x)|r;k_{1},...,k_{n-1}\right\rangle ,
\end{align}
where the electromagnetic current is%
\begin{equation}
j^{\mu}(x)=ie\phi^{+}(x)\overleftrightarrow{\partial_{\mu}}\phi(x)-2e^{2}%
A_{\mu}\phi^{+}\phi,
\end{equation}
here we assume the second term does not contribute in the infared behaviour of
the propagator and the usual commutation relation for the first term%
\begin{equation}
\delta(x_{0}-y_{0})[j_{0}(x),\phi(y)]=e\phi(x).
\end{equation}
Thus we omitt the second term in the present calculus.From the definition
(8),$T_{n}^{\mu}$ is seen to satisfy the Ward-Identity%
\begin{equation}
k_{n\mu}T_{n}^{\mu}=eT_{n-1}(r;k_{1},..,k_{n-1}).
\end{equation}
Here we mention the simple proof:%
\begin{align}
k_{n\mu}T_{n}^{\mu}  &  =\int d^{3}xk_{n\mu}\epsilon_{n}^{\mu}\exp(ik_{n}\cdot
x)\left\langle \Omega|T\phi j^{\mu}(x)|r;k_{1},...,k_{n-1}\right\rangle
\nonumber\\
&  =i\int d^{3}x\epsilon_{n}^{\mu}\exp(ik_{n}\cdot x)\partial_{\mu
}\left\langle \Omega|T\phi j^{\mu}(x)|r;k_{1},...,k_{n-1}\right\rangle
\nonumber\\
&  =e\epsilon_{n}^{\mu}\left\langle \Omega|\phi|r;k_{1},..,k_{n-1}%
\right\rangle ,
\end{align}
provided%
\begin{equation}
\partial_{\mu}T(\phi(y)j^{\mu}(x))=\delta(x-y)e\phi(y),
\end{equation}
\qquad\qquad where $e$ is the charge carried by meson.To get the solution of
the Ward-Identity we seek the non-singular contribution of photon for
$T_{n}(r;k_{1},..,k_{n}).$There is a possible solution which satisfy
Ward-Identity[9]%
\begin{equation}
T_{n}^{\mu}(r;k_{1}..k_{n})=\frac{e(2r+k_{n})^{\mu}}{2r\cdot k_{n}+k_{n}^{2}%
}T_{n-1}(r,k_{1}...k_{n-1}),r^{2}=m^{2}.
\end{equation}
Here we assume most singular contributions of photons emitted from external
lines as in four dimension.These arise from Feynman diagram in which the
soft-photon line with momentum $k_{n}^{\mu}$ and the incoming meson line can
be separated from the remainder of the diagram by cutting a single meson
line.Detailed dicussion and evaluation of the explicit form of $T_{n}$ are
given in ref [9].Of course for $n=1$%
\begin{equation}
T_{1}=\frac{e(2r+k)\cdot\epsilon}{(2r\cdot k+k^{2})},
\end{equation}
and we can determine $T_{n}$ recursively with (14).In this case $T_{n}$
becomes
\begin{equation}
T_{n}=\left(  \frac{e(2r+k)\cdot\epsilon}{2r\cdot k+k^{2}}\right)  _{n}.
\end{equation}
The one photon matrix element by LSZ is%
\begin{align}
T_{1}  &  =\left\langle in|T(\phi_{in}(x),ie\int d^{3}y\phi_{in}%
^{+}\overleftrightarrow{\partial_{\mu}^{y}}\phi_{in}(y)A_{in}^{\mu
}(y))|r;k\text{ }in\right\rangle \nonumber\\
&  =ie\int d^{3}yd^{3}z\triangle_{F}(x-y)\overleftrightarrow{\partial_{\mu
}^{y}}\epsilon_{\mu}^{\lambda}(k)\exp(ik\cdot y)\delta^{(3)}(y-z)\exp(ir\cdot
z)\nonumber\\
&  =e\frac{(2r+k)_{\mu}\epsilon_{\mu}^{\lambda}(k)}{(r+k)^{2}-m^{2}}%
\exp(i(r+k)\cdot x).
\end{align}
From this lowest order matrix element the function in (4) is given%
\begin{align}
F  &  =\sum_{one\text{ }photon}\left\langle \Omega|\phi(x)|r;k\right\rangle
\left\langle r;k|\phi^{+}(0)|\Omega\right\rangle \nonumber\\
&  =\int\frac{d^{3}k}{(2\pi)^{2}}\exp(ik\cdot x)\delta(k^{2})\theta
(k^{0})[\frac{e^{2}(2r+k)^{\mu}(2r+k)^{\nu}\Pi_{\mu\nu}}{(2r\cdot k+k^{2}%
)^{2}}],
\end{align}
and $\sigma$ is expressed by
\begin{equation}
\sigma(p)=\int\frac{d^{3}x}{(2\pi)^{3}}\exp(ip\cdot x)\int\frac{d^{2}r}%
{2r^{0}}\exp(ir\cdot x)\exp(F).
\end{equation}
Here $\Pi_{\mu\nu}$ is a polarization sum we have%
\begin{equation}
\Pi_{\mu\nu}=-(g_{\mu\nu}-\frac{k_{\mu}k_{\nu}}{k^{2}})-d\frac{k_{\mu}k_{\nu}%
}{k^{2}},
\end{equation}
and the free photon propagator%
\begin{equation}
D_{0}^{\mu\nu}=\frac{1}{k^{2}+i\epsilon}[g^{\mu\nu}-\frac{k^{\mu}k^{\nu}%
}{k^{2}}+d\frac{k^{\mu}k^{\nu}}{k^{2}}].
\end{equation}
we get%
\begin{align}
F  &  =-e^{2}\int\frac{d^{3}k}{(2\pi)^{2}}\exp(ik\cdot x)\theta(k^{0}%
)\nonumber\\
&  \times\lbrack\frac{(2r+k)^{2}\delta(k^{2})}{(2r\cdot k+k^{2})^{2}%
}+(d-1)\frac{\delta(k^{2})}{k^{2}}].
\end{align}
It is natural to set $\delta(k^{2})/k^{2}$ equals to $-\delta^{^{\prime}%
}(k^{2}).$We see the reason in calculating the one loop self energy of meson
diagram,and take imaginary part.This expression is the same as the one given
above with the interpretation $\delta(k^{2})/k^{2}=-\delta^{^{\prime}}%
(k^{2}).$We must introduce a small photon mass $\mu$ as an infrared cut
off.Therefore ( 22\ ) becomes
\begin{equation}
F=-e^{2}\int\frac{d^{3}k}{(2\pi)^{2}}\theta(k^{0})\exp(ik\cdot x)[\delta
(k^{2}-\mu^{2})(\frac{m^{2}}{(r\cdot k)^{2}}+\frac{1}{(r\cdot k)}%
)-(d-1)\frac{\partial}{\partial k^{2}}\delta(k^{2}-\mu^{2})].{}%
\end{equation}
The evaluation of the integral to $F$ is described in Appendix.The result is%
\begin{equation}
F=\frac{e^{2}}{8\pi\mu}(d-2)+\frac{\gamma e^{2}}{8\pi r}+\frac{e^{2}}{8\pi
r}\ln(\mu x)-\frac{e^{2}}{8\pi}x\ln(\mu x)-\frac{e^{2}}{8\pi}x(d+\gamma-2).
\end{equation}
It is helpful to use position space that it shows us easily the short and
long-distance behavior.The relation between perturbative spectral function and
$F(x)$ is
\begin{equation}
\sigma^{(2)}(p)=\int\frac{d^{3}x}{(2\pi)^{3}}\exp(ip\cdot x)\int\frac{d^{2}%
r}{2r^{0}}\exp(ir\cdot x)F(x).
\end{equation}
Here we think about the physical meanings of each term.
\begin{align}
\sigma(x)  &  =\frac{\exp(-mx)}{4\pi x}\exp(F(x,e,\mu,m)),\\
\sigma(p)  &  =F.T.(\sigma(x))=\int\frac{d^{3}x}{(2\pi)^{3}}\exp(ip\cdot
x)\sigma(x).
\end{align}

The terms proportional to $x$ correspond to mass renormalization and others to
wave function renormalization.
\begin{align}
\sigma(x)  &  =\exp(A)\frac{\exp(-(m+B)x)}{4\pi x}(\mu x)^{-Cx+D},\\
A  &  =\frac{e^{2}}{8\pi\mu}(d-2)+\frac{\gamma e^{2}}{8\pi m},B=\frac{e^{2}%
}{8\pi}(d+\gamma-2),C=\frac{e^{2}}{8\pi},D=\frac{e^{2}}{8\pi m}.
\end{align}
From the form of $\sigma(x)$,wave function renormalization effects denoted by
constant and $D\ln(\mu x)$ modifies the dimension of the propagator as
anomalous dimension in four dimension.Mass renormalization is proportional to
$x$ and $x\ln(\mu x).$Logarithmic corrections descrive the logarithmic
infrared divergences and adds a drastic effect as a factor $(\mu x)^{-Cx}%
$.This factor $(\mu x)^{-Cx}$ is a new type of mass singularity in three
dimension.In perturbative analysis these will be shown clear. By the
definition (1),(2),(3) $\sigma(x)$ is a full puropagator.Free propagator in
three dimension is%
\begin{equation}
\sigma_{0}(x,m+B)=\frac{\exp(-(m+B)x)}{4\pi x},\sigma_{0}(p,m+B)=\frac
{1}{(m+B)^{2}+p^{2}},
\end{equation}
and its quantum correction is expressed by $\exp(A)(\mu x)^{-Cx+D}.$We find in
this expression that $\sigma(x)$ is finite provided
\begin{align}
|\sigma(x)  &  =\exp(A)\frac{\exp(-(m+B)x)}{4\pi x}(\mu x)^{-Cx+D}|=finite,\\
0  &  \leq\int_{0}^{\infty}\sigma(x)dx\leq M,\text{ }0<D,
\end{align}

and there exists $\sigma(p)$%
\begin{equation}
\sigma(p)=\int_{0}^{\infty}\frac{x\sin(px)}{p}\sigma(x)dx.
\end{equation}
In this case we have not a simple pole even in the Yennie gauge $d=2$ in which
we see the singularity at $p^{2}=(m+e^{2}\gamma/8\pi)^{2}.$Next we treat the
momentum dependece of the propagator $\sigma(p)$ (32)$.$The integral is not
analytic and we cannot get the precise expression for $\sigma(p).$

If we expand $(\mu x)^{-Cx+D}$%
\begin{equation}
(\mu x)^{-Cx+D}=1+(-Cx+D)\ln(\mu x)+\frac{1}{2}(-Cx+D)^{2}(\ln(\mu
x))^{2}+\frac{1}{6}(-Cx+D)^{3}(\ln(\mu x))^{3}+..,
\end{equation}
term by term integral of $\sigma(p)$ converges.Using the following integrals
\begin{align}
I_{1}  &  =\int_{0}^{\infty}\frac{\sin(px)}{p}\exp(-mx)\ln(\mu x)dx\nonumber\\
&  =-\frac{\gamma}{p^{2}+m^{2}}-\frac{\ln((m^{2}+p^{2})/\mu^{2})}%
{2(p^{2}+m^{2})}-\frac{\ln((m-\sqrt{-p^{2}})/(m+\sqrt{-p^{2}}))}{p^{2}+m^{2}},
\end{align}%
\begin{align}
I_{2}  &  =\int_{0}^{\infty}\frac{\sin(px)}{p}\exp(-mx)x\ln(\mu
x)dx\nonumber\\
&  =\frac{-m}{(p^{2}+m^{2})^{2}}[\ln((m-\sqrt{-p^{2}})/(m+\sqrt{-p^{2}}%
))+\ln((p^{2}+m^{2})/\mu^{2})-2(1-\gamma)],
\end{align}
$\sigma(p)$ up to $O(e^{2})$ is given%
\begin{align}
\sigma^{(2)}(p)  &  =[\frac{1}{m^{2}+p^{2}}+\frac{A}{p^{2}+m^{2}}-\frac
{mB}{2(p^{2}+m^{2})^{2}}\nonumber\\
&  +DI_{1}-CI_{2}],p=\sqrt{-p^{2}}.
\end{align}
In this case the renormalization constant $Z$ (residue of the pole term)
becomes%
\begin{equation}
Z=1+A+D(-\gamma+\frac{1}{2}\ln(\frac{p^{2}+m^{2}}{\mu^{2}}))_{p^{2}%
\rightarrow\infty}\rightarrow\infty.
\end{equation}
In the Yennie gauge there remains a renormalization%
\begin{equation}
Z=1+\frac{e^{2}}{16\pi m}\ln((p^{2}+m^{2})/\mu^{2})_{p^{2}\rightarrow\infty
}\rightarrow\infty.
\end{equation}
In Minkowski space, $p^{2}\rightarrow-p^{2},$with dicontinuity%
\begin{align}
\frac{1}{x-i\epsilon}  &  =P.V\frac{1}{x}+i\pi\delta(x)\\
\frac{1}{(x-i\epsilon)^{2}}  &  =P.V\frac{1}{x^{2}}+i\pi\delta^{^{\prime}%
}(x),\\
\frac{1}{x-i\epsilon}\ln(x-i\epsilon)  &  =i\pi\delta(x)\ln(x)+P.V\frac{1}%
{x}i\pi.
\end{align}
we can determine the the strucure near $p^{2}=-m^{2}$ by the imaginary part of
$\sigma(p).$Here we notice the second-order spectral function $\sigma
^{(2)}(p)$%
\begin{align}
\sigma^{(2)}(x)  &  =\frac{\exp(-mx)}{4\pi x}(1+A-Bx+(D-Cx)\ln(\mu x)),\\
\sigma^{(2)}(p)  &  =\frac{1}{m^{2}+p^{2}}+\frac{A-\gamma D}{m^{2}+p^{2}%
}-\frac{2mB}{(m^{2}+p^{2})^{2}}\nonumber\\
&  -D\frac{\ln(\sqrt{(m^{2}+p^{2})/\mu^{2}})}{(p^{2}+m^{2})}\nonumber\\
&  +\frac{Cm}{(p^{2}+m^{2})^{2}}[\ln((p^{2}+m^{2})/\mu^{2})-2(1-\gamma)].
\end{align}
Here we look at mass renormalization part
\begin{align}
Z_{m}  &  =\frac{m(2B+2C(1-\gamma)-C\ln((p^{2}+m^{2})/\mu^{2}))}{p^{2}+m^{2}%
}\nonumber\\
&  =\frac{e^{2}}{8\pi}\frac{2(d-1)-\ln((p^{2}+m^{2})/\mu^{2})}{p^{2}+m^{2}}.
\end{align}
Spectral function $\operatorname{Im}\sigma^{(2)}(-p^{2})$ reads%
\begin{align}
\frac{\operatorname{Im}\sigma^{(2)}(-p^{2})}{\pi}  &  =(1+A-\gamma
D)\delta(p^{2}-m^{2})-\frac{e^{2}}{4\pi}m(d-1)\delta^{^{\prime}}(p^{2}%
-m^{2})\nonumber\\
&  -\frac{D\theta(p^{2}-m^{2})}{m^{2}-p^{2}}+\frac{Cm\theta(p^{2}-m^{2}%
)}{(m^{2}-p^{2})^{2}}.
\end{align}
Usually we estimate the second-order spectral function in the Feynman gauge
$d=1;$%
\begin{align}
\frac{\operatorname{Im}\sigma^{(2)}(-p^{2})}{\pi}_{d=1}  &  =[(1-\frac{e^{2}%
}{8\pi\mu})\delta(p^{2}-m^{2})\nonumber\\
&  [\frac{D}{p^{2}-m^{2}}+\frac{Cm}{(m^{2}-p^{2})^{2}}]\theta(p^{2}-m^{2})].
\end{align}
This shows the renormalization constant $Z$ has an linear infrared divergence
as $\mu\rightarrow0$.If we add higher order corrections linear infrared
divergences cancell each other in the same mechanism in four
dimension[15,16,17]. In the Yennie gauge it has cuts
\begin{align}
\frac{\operatorname{Im}\sigma^{(2)}(-p^{2})}{\pi}_{d=2}  &  =\delta
(p^{2}-m^{2})-\frac{e^{2}m}{4\pi}\delta^{^{\prime}}(p^{2}-m^{2})\nonumber\\
&  +\frac{e^{2}}{8\pi m}(\frac{1}{p^{2}-m^{2}}+\frac{2m^{2}}{(p^{2}-m^{2}%
)^{2}})\theta(p^{2}-m^{2}).
\end{align}
To see the singularity structure near $p^{2}=-m^{\ast2}$ it is better to
expand around $p^{2}=-m^{\ast2}.$We get%
\begin{equation}
\sigma^{(2)}(p)=[\frac{1}{m^{\ast2}+p^{2}}+(\exp(A)-1)\times(-CI_{2}%
(m\rightarrow m^{\ast})+DI_{1}(m\rightarrow m^{\ast})].
\end{equation}
In this case the renormalization constant $Z:$
\begin{equation}
Z=1+(\exp(A)-1)+D(-\gamma+\frac{1}{2}\ln(\frac{m^{\ast2}+p^{2}}{\mu^{2}}))
\end{equation}
is divergent at $p^{2}=-m^{\ast2}.$There is an interesting contribution from
$\exp(F_{2}{\normalsize )}$%
\begin{align}
&  \int_{0}^{\infty}\frac{\sin(px)}{p}\exp(-mx)(\mu x)^{a}dx\nonumber\\
&  =-\Gamma(a+1)\frac{\cos(\pi a)}{2}(p^{2}+m^{2})^{-1-a/2}\mu^{a}\nonumber\\
&  \times\frac{1}{\sqrt{-p^{2}}}[(\sqrt{-p^{2}}+m)(\frac{\sqrt{-p^{2}}%
-m}{\sqrt{-p^{2}}+m})^{-a/2}+(\sqrt{-p^{2}}-m)(\frac{\sqrt{-p^{2}}+m}%
{\sqrt{-p^{2}}-m})^{a/2}]\nonumber\\
&  \sim(\sqrt{-p^{2}}-m)^{-1-a}\text{ near }p^{2}=-m^{2},a=\frac{e^{2}}{8\pi
m}.
\end{align}

\bigskip Here we notice that this type of singularity also appears in four
dimension where $a=-\alpha(d-3)/(2\pi)$ but in this case $a$ is gauge
invariant.Formally fourier transform of $\sigma(p)$ can be written by double
fourier transform%
\begin{equation}
\sigma(p)=\int\frac{d^{3}q}{(2\pi)^{3}}F.T.(\frac{\exp(-mx)}{x}(\mu
x)^{D})(q)\times F.T.(\mu x)^{-Cx}(p-q),
\end{equation}
but we do not discuss details of it in this paper.As we see in the annalysis
in terms of $e^{2}$ ,the contribution of the function $(\mu x)^{-Cx}$ leads to
a new type of mass singularity ;divergent series at $p^{2}=-m^{\ast2}$ in
$\sigma^{(2)}(p)$.We have seen that the fourier transformation of the
propagator is very difficult.It is interesting to study the phase structure of
the model.First we estimate renormalization constant by the following sum
rule[10,11],%
\begin{equation}
Z^{-1}=\int\sigma(\omega)d\omega,
\end{equation}%
\begin{equation}
Z^{-1}=\lim_{p^{2}\rightarrow\infty}p^{2}\int\sigma(\omega)d\omega\frac
{\sin(px)}{p}\frac{\exp(-\omega x)}{4\pi x}dx,
\end{equation}%
\begin{equation}
\lim_{p\rightarrow\infty}F(p)=\lim_{p\rightarrow\infty}\int_{0}^{\infty}%
\frac{\pi x\sin(px)}{p}F(x)dx=\lim_{p\rightarrow\infty}\frac{\pi}{p^{2}%
}(xF(x))(0),
\end{equation}
we can evaluate this quantity by direct substitution of $\sigma(x)$ into the
above equations and take the limit
\begin{align}
Z^{-1}  &  =\exp(A)\lim_{x\rightarrow0_{+}}\exp(-mx)(\mu x)^{-Cx+D}\nonumber\\
&  =0.
\end{align}

Next we consider the gauge dependence of the propagator[12,13,14].When we
write the photon propagator as%
\begin{equation}
D_{\mu\nu}(k)=D_{\mu\nu}^{(0)}(k)+k_{\mu}k_{\nu}M(k),
\end{equation}
we seek the change of the propagator.Under the gauge transformation defined:%
\begin{align}
\phi(x)  &  \rightarrow\exp(ie\chi(x)),\phi^{+}(x)\rightarrow\phi^{+}%
(x)\exp(-ie\chi(x)),\nonumber\\
A_{\mu}(x)  &  \rightarrow A_{\mu}(x)-\partial_{\mu}\chi,\delta(\phi
^{+}\overleftrightarrow{\partial_{\mu}}\phi)=-\partial_{\mu}\chi(\phi
^{+}\overleftrightarrow{\partial_{\mu}}\phi-2ie\phi^{+}\phi A_{\mu}),
\end{align}
propagator changes
\begin{align}
D_{\mu\nu}  &  \rightarrow i\left\langle T[A_{\mu}(x)-\partial_{\mu}^{x}%
\chi(x)][A_{\nu}(y)-\partial_{\nu}^{y}(y)]\right\rangle \nonumber\\
&  =D_{\mu\nu}+i\left\langle T\partial_{\mu}^{x}\chi(x)\partial_{\nu}%
^{y}(y)\right\rangle ,\\
\delta D_{\mu\nu}  &  =\partial_{\mu}^{x}\partial_{\nu}^{y}%
M(x-y),M(x-y)=i\left\langle T\chi(x)\chi(y)\right\rangle \nonumber\\
\Delta &  \rightarrow-i\left\langle T\phi(x)\exp(ie\chi(x)\exp(-ie\chi
(y)\phi^{+}(y)\right\rangle \nonumber\\
=  &  \Delta\left\langle T\exp(ie\chi(x))\exp(-ie\chi(y))\right\rangle .
\end{align}
Here we expand in $\delta\chi$%
\begin{align}
\left\langle T\exp(ie\delta\chi(x))\exp(-ie\delta\chi(y))\right\rangle  &
=-\frac{1}{2}e^{2}\left\langle \delta\chi^{2}(x)+\delta\chi^{2}(y)-2T\delta
\chi(x)\delta\chi(y)\right\rangle ,\nonumber\\
\delta\Delta &  =ie^{2}\Delta^{(0)}(x-y)[M(0)-M(x-y)],\nonumber\\
i[M(0)-M(x-y)]  &  =-\left\langle T\delta\chi(x)\delta\chi(y)\right\rangle .
\end{align}
Usually $M$ is identified as the gauge fixing term%
\begin{align}
M(k)  &  =-d/k^{4},\nonumber\\
ie^{2}M(x)  &  =-ie^{2}d\int\frac{d^{3}k}{(2\pi)^{3}}\frac{\exp(ik\cdot
x)-1}{k^{4}}\nonumber\\
&  =-\frac{e^{2}d}{8\pi}x,
\end{align}
with subtraction of infrared divergence at $k=0.$ We get the propagator(59) in
covariant $d$ gauge for three dimension%
\begin{align}
\Delta(d,x)  &  =\exp(-\frac{e^{2}d}{8\pi}x)\Delta(0,x)\\
&  =\exp(-\frac{e^{2}d}{8\pi}x)\sigma(0,x),
\end{align}
Since $\sigma(x)$ is a full propagator it obeys the Landau-Kharatonikov
transformation in our approximation except for linear infrared divergent
factor $A$%
\begin{equation}
\sigma(d,x)=\frac{\exp(-(m^{\ast}+de^{2}/8\pi)x)}{x}\exp(A)(\mu x)^{-Cx+D}.
\end{equation}
Therefore we notice that the Landau-Karatonikov transformation is
regularization dependent.We keep the Landau gauge and not take the covariant
$d$ gauge to avoid longitudinal photon.Here we mention the radiative
correction of mass to understand the mass shift$.$In the Landau gauge our
approximation changes the mass $m$ to $m^{\ast}=m+e^{2}(\gamma-2)/(8\pi).$The
constant $\gamma$ appeared after the expansion of $\operatorname{Ei}(1,\mu x)$
in $\mu.$In ref[5],gauge dependent self-energy is estimated at $O(e^{2})$%
\begin{equation}
\Sigma(m)=d\frac{e^{2}}{4\pi},
\end{equation}
and the physical mass for fermion defined in the Landau gauge
\begin{equation}
m_{phy}=m+\Sigma(m).
\end{equation}
If we estimate one-loop self-energy in the Landau gauge
\begin{equation}
\Sigma(m)=\frac{e^{2}}{2\pi},
\end{equation}
which is two-times larger than $e^{2}/(4\pi)$ in $m^{\ast}$ since we used
$D_{+}(x)$ instead of $D(x).$We see the same effects occured for scalar with
vertex correction.We call $m^{\ast}=m+e^{2}(\gamma-2)/8\pi$ the renormalized
mass in our approximation.At first sight, mass shift seems to be peculiar but
infrared behaviour is governed by one-particle intermediate state.

\section{Spinor QED}

In this section we study the spectral function for massive fermion.Similar to
scalar case propagator and the spectral function are defined%
\begin{align}
S(p)  &  =\int d^{3}x\exp(ip\cdot x)\left\langle \Omega|T(\psi(x)\overline
{\psi}(0))|\Omega\right\rangle \nonumber\\
&  =(\int_{m}^{\infty}+\int_{-\infty}^{-m})\frac{\rho(\omega)d\omega}%
{p\cdot\gamma-\omega+i\epsilon},\\
\rho(p^{2})  &  =(2\pi)^{2}\sum_{N}\delta^{(3)}(p-p_{N})\left\langle
\Omega|\psi|N\right\rangle \left\langle N|\overline{\psi}|\Omega\right\rangle
.
\end{align}

Matrix element is%
\begin{align}
T_{n}  &  =\left\langle \Omega|\psi|r;k_{1},..,k_{n}\right\rangle ,\\
T_{n}^{\mu}  &  =-\int d^{3}x\exp(ik_{n}\cdot x)\left\langle \Omega|T\psi
j^{\mu}|r;k_{1},...k_{n-1}\right\rangle ,
\end{align}
provided%
\begin{equation}
\square_{x}T\psi A_{\mu}(x)=T\psi\square_{x}A_{\mu}(x)=T\psi(-j_{\mu
}(x)+\partial_{\mu}^{x}(\partial\cdot A(x))).
\end{equation}
In the similar way to the scalar case $T_{n}$ satisfies Ward-Identity:
\begin{align}
\partial_{\mu}^{x}T(\psi j_{\mu}(x))  &  =-e\psi(x),\\
\partial_{\mu}^{x}T(\overline{\psi}j_{\mu}(x))  &  =e\overline{\psi}(x),\\
k_{n\mu}T_{n}^{\mu}(r,k_{1},k_{2},..k_{n})  &  =eT_{n-1}(r,k_{1}%
,k_{2},..k_{n-1}),r^{2}=m^{2}.
\end{align}
One photon matrix element by LSZ is
\begin{align}
T_{1}  &  =\left\langle in|T(\psi_{in}(x),ie\int d^{3}y\overline{\psi}%
_{in}(y)\gamma_{\mu}\psi_{in}(y)A_{in}^{\mu}(y))|r;k\text{ }in\right\rangle
\nonumber\\
&  =ie\int d^{3}yd^{3}zS_{F}(x-y)\gamma_{\mu}\delta^{(3)}(y-z)\exp(i(k\cdot
y+r\cdot z))\epsilon^{\mu}(k,\lambda)U(r,s)\nonumber\\
&  =-ie\frac{(r+k)\cdot\gamma+m}{((r+k)^{2}-m^{2})}\gamma_{\mu}\epsilon^{\mu
}(k,\lambda)\exp(i((k+r)\cdot x))U(r,s),
\end{align}
where $U(r,s)$ is a two-component free particle spinor with positive energy:%
\begin{equation}
\sum_{S}U(r,s)\overline{U}(r,s)=\frac{\gamma\cdot r+m}{2m}.
\end{equation}
In this case the function $F$ becomes%
\begin{align}
F  &  =e^{2}\int\frac{d^{3}k}{(2\pi)^{3}}\exp(ik\cdot x)\theta(k^{0}%
)\delta(k^{2})tr[\frac{(r+k)\cdot\gamma+m}{((r+k)^{2}-m^{2})}\gamma_{\mu}%
\frac{\gamma\cdot r+m}{2m}\gamma_{\nu}\frac{(r+k)\cdot\gamma+m}{((r+k)^{2}%
-m^{2})}\Pi^{\mu\nu}]\nonumber\\
&  =-e^{2}\int\frac{d^{3}k}{(2\pi)^{3}}\exp(ik\cdot x)\theta(k^{0}%
)[\delta(k^{2}-\mu^{2})(\frac{m^{2}}{(r\cdot k)^{2}}+\frac{1}{(r\cdot
k)})+(d-1)\frac{\partial\delta(k^{2}-\mu^{2})}{\partial k^{2}}].
\end{align}%
\begin{equation}
F=\frac{e^{2}m^{2}}{8\pi r^{2}}(-\frac{1}{\mu}+x(1-\gamma)-x\ln(\mu
x))+\frac{e^{2}}{8\pi r}(\ln(\mu x)+\gamma)+\frac{e^{2}}{8\pi}(d-1)(\frac
{1}{\mu}-x).
\end{equation}
Thus the lowest order spectral function is exactly the same with that in the
scalar case except for the normalization factor in the phase space integral:%

\begin{align}
\rho(p^{2})  &  =\int\frac{d^{3}x}{(2\pi)^{3}}\exp(ip\cdot x)\rho
(x)\nonumber\\
&  =\int\frac{d^{3}x}{(2\pi)^{3}}\exp(ip\cdot x)\int d^{2}r\frac{m}{r_{0}}%
\exp(ir\cdot x)\exp(F).
\end{align}
After angular integral we get
\begin{equation}
\rho(p)=\frac{m}{4\pi p}\exp(A)\int_{0}^{\infty}dx\sin(px)\frac{\exp
(-(B+m)x)}{x}(\mu x)^{-Cx+D}.
\end{equation}
Therefore the structure is the same with scalar case as we mentioned in the
last section. Consequently there is no infrared divergences at $p^{2}=m^{2}.$

For the renormalization constants we apply the same argument as for the
scalar,full propagator is expressed in the dispersion integral%
\[
S_{F}(x)=-\int\rho(\omega)d\omega(i\gamma\cdot\partial+\omega)\frac
{\exp(-\omega x)}{4\pi x}=-(i\gamma\cdot\partial\sigma(x)+\sigma(x)),
\]
Here we use the fourier transform of $S_{F}$ to determine $Z_{2}^{-1}.$For
free case the propagator in position space is%
\begin{equation}
S_{F}^{(0)}(x)=(i\gamma\cdot\partial)\frac{1}{4\pi\sqrt{-x^{2}}}=\frac
{i\gamma\cdot x}{4\pi(-x^{2})^{3/2}}=-\frac{i\gamma\cdot x}{\sqrt{-x^{2}}%
}\frac{1}{4\pi x^{2}}.
\end{equation}
We see the dimension of the $S_{F}^{(0)}(x)$ is equal to $1/x^{2}.$I momentum
space we obtain%
\begin{equation}
\int_{0}^{\infty}\frac{x\sin(px)}{p}\frac{1}{x^{2}}dx=\frac{\pi}{2p},
\end{equation}
and this shows the ordinary expression in momentum space%
\begin{equation}
S_{F}^{(0)}(p)=\frac{\gamma\cdot p}{p^{2}}=\frac{\gamma\cdot p}{p}\frac{1}{p}.
\end{equation}%
\begin{align}
Z_{2}^{-1}  &  =\lim_{x\rightarrow0_{+}}[(i\gamma\cdot\partial\sigma
(\sqrt{-x^{2}}))\frac{i\gamma\cdot x}{\sqrt{-x^{2}}}]=\lim_{x\rightarrow0_{+}%
}(\frac{d\sigma(\sqrt{-x^{2}})}{d\sqrt{-x^{2}}})\\
&  =\pi\exp(A)\lim_{x\rightarrow0_{+}}\frac{d}{dr}(\frac{\exp(-mr)}{r}(\mu
r)^{-Cr+D})=\left[
\begin{array}
[c]{cc}%
0 & (1<D)\\
\infty & (1\geq D)
\end{array}
\right]  ,\\
\frac{d}{dr}(\frac{\exp(-mr)}{r}(\mu r)^{-Cr+D})  &  =\exp(-mr)(\mu
r)^{-Cr+D}[-\frac{m+C}{r}+\frac{D-1}{r^{2}}-\frac{C\ln(\mu r)}{r}]\nonumber\\
,r  &  =\sqrt{-x^{2}}.\\
m_{0}Z_{2}^{-1}  &  =\int\omega\rho(\omega)d\omega=\exp(A)\lim_{x\rightarrow
0_{+}}(\mu x)^{-Cx+D}=0.
\end{align}
Therefore there is a confining phase;$Z_{2}=0(Z_{2}^{-1}=\infty)$ of charged
particle for weak coupling constant$.$In this case bare mass vanishes for all
coupling. Order parameter for the vacuum expectation value of pair condensate
is given
\begin{align}
\left\langle \overline{\psi}\psi\right\rangle  &  =-itrS_{F}(x)=-2m^{2}%
\exp(A)\lim_{x\rightarrow0_{+}}(\exp(-mx)(x)^{-Cx+D-1}))\nonumber\\
&  =-2m^{2}\left[
\begin{array}
[c]{cc}%
0 & (1<D)\\
finite & (1=D)\\
\infty & (1>D)
\end{array}
\right]  .
\end{align}
Here we notice that there is a critical coupling constant $D_{cr}%
=1(e^{2}/(8\pi m)=1).$

\section{Summary}

We have seen how the Ward-Identity for soft photon may be applied to
three-dimensional electro-dynamics to extract full propagator in position
space and the infrared behaviour of the propagator in momentum
space.Neglecting unconventional terms in the elecrtomagnetic current for
scalar,spectral functions for scalar and spinor coincide each
other.Exponentiation of the lowest order spectral function corresponds to the
infinite ladder approximation. Since the theory is super renormalizable the
mass are corrected to add some finite radiative correction.We found a new type
of mass singularity in three dimensional QED in position space as $(\mu
x)^{-Cx}.$ In momentum space it shows us logarithmic mass renormalization and
yields a singular infrared structure of the propagator. Wave functuin
renormalization $D$ plays the role as anoumalous dimension .These are the
consequences of logarithmic infrared divergences. Renormalization constant
vanishes for spinor case and there is a confining phase for weak coupling.This
picture is consistent with perturbative logarithmic infrared divergences at
$p^{2}=m^{2}.$If the coupling constant is smaller than $D_{cr}$ ($D=e^{2}/8\pi
m\leq1),$vacuum expectation value $\left\langle \overline{\psi}\psi
\right\rangle $ becomes finite.In our lowest order spectral function there
remains linear infrared divergences that was regularized by photon
mass.$O(e^{2})$ correction to the external line and the cancellation of
infrared divergences are now in progress[17].

\section{Aknowledgement}

The author would like thank Prof.Roman Jackiw to his introduction on technique
based on Ward-Identity at MIT September 2002.

\section{Appendix}

{\normalsize To evaluate }$F$ we use {\normalsize the function }$D_{+}(x)$ in
three dimension:%

\begin{align}
D_{+}(x)  &  =\frac{1}{(2\pi)^{2}i}\int\exp(ik\cdot x)\theta(k^{0}%
)\delta(k^{2}-\mu^{2})d^{3}k\nonumber\\
&  =\frac{1}{(2\pi)^{2}i}\int_{0}^{\infty}J_{0}(kx)\frac{\pi kdk}{2\sqrt
{k^{2}+\mu^{2}}}=\frac{\exp(-\mu x)}{8\pi ix},
\end{align}
{\huge \qquad}

with the following parameter trick as in ref[9]%

\begin{align}
\lim_{\epsilon\rightarrow0_{+}}\int_{0}^{\infty}d\alpha\exp(i(k+i\epsilon
)\cdot(x+\alpha r))  &  =\frac{\exp(ik\cdot x)}{k\cdot r},\\
\lim_{\epsilon\rightarrow0_{+}}\int_{0}^{\infty}\alpha d\alpha\exp
(i(k+i\epsilon)\cdot(x+\alpha r))  &  =\frac{\exp(ik\cdot x)}{(k\cdot r)^{2}}.
\end{align}
The function $F$ is written in terms of the parameter integrals
\begin{align}
F  &  =ie^{2}m^{2}\int_{0}^{\infty}\alpha d\alpha D_{+}(x+\alpha r,\mu
)-e^{2}\int_{0}^{\infty}d\alpha D_{+}(x+\alpha r,\mu)\nonumber\\
&  -ie^{2}(d-1)\frac{\partial}{\partial\mu^{2}}D_{+}(x,\mu)\nonumber\\
&  =\frac{e^{2}m^{2}}{8\pi r^{2}}(-\frac{\exp(-\mu x)}{\mu}+x\operatorname{Ei}%
(1,\mu x))-\frac{e^{2}}{8\pi r}\operatorname{Ei}(1,\mu x)+(d-1)\frac{e^{2}%
}{8\pi\mu}\exp(-\mu x),
\end{align}
where the function $\operatorname{Ei}(n,x)$ is defined
\begin{equation}
\operatorname{Ei}(n,x)=\int_{1}^{\infty}\frac{\exp(-xt)}{t^{n}}dt.
\end{equation}
It is understood that all terms which vanish with $\mu\rightarrow0$ are
ignored.
\begin{equation}
\operatorname{Ei}(1,\mu x)=-\gamma-\ln(\mu x)+O(\mu x),
\end{equation}%
\begin{align}
F_{1}  &  =\frac{e^{2}m^{2}}{8\pi r^{2}}(-\frac{1}{\mu}+x(1-\ln(\mu
x)-\gamma))+O(\mu),\\
F_{2}  &  =\frac{e^{2}}{8\pi r}(\ln(\mu x)+\gamma)+O(\mu),\\
F_{g}  &  =\frac{e^{2}}{8\pi}(\frac{1}{\mu}-x)(d-1)+O(\mu).
\end{align}
Here $\gamma$ is Euler's constant.Using the integrals%
\begin{equation}
\int d^{3}x\exp(ip\cdot x)\int d^{3}r\delta(r^{2}-m^{2})\exp(ir\cdot
x)f(r)=f(m),
\end{equation}%
\begin{equation}
\int d^{3}x\exp(ip\cdot x)\int\frac{d^{3}r}{(2\pi)^{3}}\delta(r^{2}%
-m^{2})=\frac{1}{m^{2}+p^{2}}.
\end{equation}
we get $F$ in position space%
\begin{equation}
F=\frac{e^{2}}{8\pi\mu}(d-2)+\frac{\gamma e^{2}}{8\pi r}+\frac{e^{2}}{8\pi
r}\ln(\mu x)-\frac{e^{2}}{8\pi}x\ln(\mu x)-\frac{e^{2}}{8\pi}x(d+\gamma-2).
\end{equation}
After integration over $r$
\begin{align}
\int_{0}^{\pi}\exp(ir\cdot x\cos(\theta))d\theta &  =\pi J_{0}(rx),\\
\int_{0}^{\infty}dr\frac{\pi rJ_{0}(rx)}{\sqrt{r^{2}+m^{2}}}  &  =\frac
{\exp(-mx)}{x},
\end{align}
we obtain the full propagator%
\begin{align}
\sigma(x)  &  =\exp(A)\frac{\exp(-(m+B)x)}{4\pi x}(\mu x)^{-Cx+D},\\
A  &  =\frac{e^{2}}{8\pi\mu}(d-2)+\frac{\gamma e^{2}}{8\pi m},B=\frac{e^{2}%
}{8\pi}(d+\gamma-2),C=\frac{e^{2}}{8\pi},D=\frac{e^{2}}{8\pi m}.
\end{align}

\section{References}

[1] R.Jackiw,S.Templeton,Phys.Rev.\textbf{23D}.2291(1981)

[2] E.I.Guendelman,Z.M.Raulvic,Phys.Rev.\textbf{30D}.1338(1984)

[3] T.Appelquist,D,Nash,L.C.R.Wijewardhana,Phys.Rev.Lett.\textbf{60}.2575(1988)

[4] Y.Hoshino,T.Matsuyama,Phys.Lett.\textbf{222B}.493(1989)

[5] S.Deser,R.Jackiw,S.Templeton,Ann.Phys.\textbf{140}.372(1982)

[6] A.B.Waites,R.Delbourgo,Int.J.Mod.Phys.\textbf{27A}.6857(1992)

[7] D.Atkinson,D.W.E.Blatt,Nucl.Phys.\textbf{151B}.342(1979)

[8] P.Maris,Phys.Rev.\textbf{52D}.6087(1995);Y.Hoshino,IL Nuovo
Cim.\textbf{112A}.335(1999)

[9] R.Jackiw,L.Soloviev,Phys.Rev\textbf{173}.1458(1968)

[10] R.Delbourgo,IL Nuovo Cim.\textbf{49A}.485(1978)

[11]K.Nishijima,Prog.Theor.Phys;\textbf{81}.878(1989),ibid.\textbf{83}.1200(1990)

[12]L.D.Landau,J,M.Kharatonikov,Zk.Eksp.Theor.Fiz.\textbf{29}.89(1958)

[13]B.Zumino,J.Math,Phys.\textbf{1}.1(1960)

[14] in Landau,Lifzhits,Quantum Electrodynamics (Butterworth Heinemann)

[15]T.Kinoshita,Prog.Theor.Phys.\textbf{5},1045(1950)

[16]N.Nakanishi,Prog.Theor.Phys.\textbf{19},159(1958)

[17] Y.Hoshino,in preparation

\end{document}